\documentstyle[12pt,aasms4,psfig]{article}

\slugcomment{to appear in the Astrophysical Journal (Letters)} 
 
\begin{document}
 
\title{Dust Emission from High Redshift QSOs}

 
\author{
C. L. Carilli$^{1,2}$,
F. Bertoldi$^1$,
K.M. Menten$^1$,
M.P. Rupen$^2$,
E. Kreysa$^1$,
Xiaohui Fan$^3$,
Michael A. Strauss$^3$,
Donald P. Schneider$^4$,
A. Bertarini$^1$,
M.S. Yun$^2$,
R. Zylka$^1$
}          

\affil{$^{1}$Max-Planck-Institut f\"{u}r Radioastronomie,
Auf dem H\"ugel 69, D-53121 Bonn, Germany}

\affil{$^{2}$National Radio Astronomy Observatory, P.O. Box O,
Socorro, NM 87801, USA }

\affil{$^{3}$Princeton University Observatory, Peyton Hall, Princeton,
NJ 08544, USA}

\affil{$^{4}$Dept. of Astronomy, Pennsylvania State University, 
University Park, PA 16802, USA} 

\vskip 0.2in
\affil{ccarilli@nrao.edu}      

\begin{abstract}

We present detections of emission at 250 GHz (1.2 mm) from two high
redshift QSOs from the Sloan Digital Sky Survey sample using the
bolometer array at the IRAM 30m telescope. The sources are 
SDSSp 015048.83+004126.2
at $z = 3.7$, and SDSSp J033829.31+002156.3
at $z = 5.0$, which 
is the third highest redshift QSO known, and the highest
redshift mm emitting source yet identified.  We also present deep
radio continuum imaging of these two sources at 1.4 GHz using the Very
Large Array.  The combination of cm and mm observations indicate that
the 250 GHz emission is most likely thermal dust emission, with
implied dust masses $\approx 10^8$ M$_\odot$.  We consider possible
dust heating mechanisms, including UV emission from the active nucleus
(AGN), and a massive starburst concurrent with the AGN, with implied
star formation rates $> 10^3$ M$_\odot$ year$^{-1}$.

\end{abstract}
 

\section {Introduction}

Modern telescopes operating from radio through optical wavelengths are
detecting star forming galaxies out to redshifts $z > 4$ (Steidel et
al. 1999, Bunker \& van Breugel 2000, Adelberger \& Steidel 2000).  
These observations are pushing into the `dark
ages,' the epoch when the first stars and/or black holes may have
formed (Rees 1999).  Millimeter (mm) and submm
observations provide a powerful probe into this era due to the
sharp rise of observed flux density with increasing frequency in the
modified Rayleigh-Jeans portion of the grey-body spectrum for thermal
dust emission from galaxies.  Millimeter and submm surveys thereby
provide a uniquely {\it distance independent} sample of objects in the
universe for $z > 0.5$ (Blain \& Longair 1993). These surveys have
revealed a population of dusty, luminous star forming galaxies
at high redshift which may correspond to forming spheroidal galaxies
(Smail, Ivison, \& Blain 1998, Hughes et al. 1998, Barger et
al. 1998, Eales et al. 1998, Bertoldi et al. 2000).  An interesting
sub-sample of dust emitting sources at high redshift are active
galaxies, including powerful radio galaxies (Chini \& Kr\"ugel 1994,
Hughes \& Dunlop 1999, Cimatti et al. 1999, Best et al. 1999,
Papadopoulos et al. 1999, Carilli et al. 2000), and optically selected
QSOs (Omont et al. 1996a).

In an extensive survey at 240 GHz, Chini, 
Kreysa, \& Biermann (1989) found that the majority of $z < 1$ QSOs
show dust emission with dust masses $\approx$ few$\times 10^{7}$
M$_\odot$, comparable to normal spiral galaxies. 
They argue that the dominant dust heating mechanism is
radiation from the active nucleus (AGN), based primarily on spectral
indices between mm and X-ray wavelengths. On the other hand, Sanders
et al. (1989) showed that the majority of radio quiet QSOs in the PG
sample show spectral energy distributions from cm to submm wavelengths
consistent with star forming galaxies. However, they suggest that 
this  may be coincidental, since  concurrent starbursts would
require large star formation rates to power the dust emission, while
it would require the
absorption of only a fraction of the AGN UV luminosity by dust.

Omont et al. (1996a) extended the 240 GHz study of QSOs to high
redshift by observing a sample of  $z > 4$ QSOs from the Automatic
Plate Measuring (APM) survey.  They found that 6 of 16 sources show
dust emission at 3 mJy or greater, with implied FIR luminosities
$\ge$ 10$^{13}$ L$_\odot$, and dust masses $\ge 10^8$ M$_\odot$.
Follow-up observations of three of these dust-emitting QSOs revealed
CO emission as well, with implied molecular gas masses $\approx$ few
$10^{10}$ M$_\odot$ (Guilloteau et al. 1997, 1999, Ohta et al. 1996,
Omont et al. 1996b, Carilli, Menten, \& Yun 1999).  
Given the large dust and gas
masses, Omont et al. (1996a) made the circumstantial argument that the
dominant dust heating mechanism may be star formation. Supporting
evidence came from deep radio observations at 1.4 GHz, which showed
that the ratio of the radio continuum to submm continuum emission from
these sources is consistent with the well established 
radio-to-far IR 
correlation for low redshift star forming galaxies (Yun et al. 1999).
All these data
(dust, CO, radio continuum) suggest that the host galaxies of these
QSOs are gas-rich, and may be forming stars at a rate $\ge$ 10$^3$
M$_\odot$ year$^{-1}$, although it remains unclear to what extent the
AGN plays a role in heating the dust and powering the radio emission.

We have begun an extensive observational program 
at cm and mm wavelengths on a sample of high
redshift QSOs from the Sloan Digital Sky Survey (SDSS; York
et al. 2000).  The sample is the result of optical 
spectroscopy of objects of
unusual color from  the northern Galactic Cap and the Southern
Equatorial Stripe, which  has yielded 40 QSOs with $z \ge 3.6$, 
including the four highest redshift QSOs known
(Fan et al. 1999, Fan et al. 2000,
Schneider et al. 2000).  This sample presents an ideal opportunity to
investigate the properties of the most distant QSOs and their host
galaxies.  The SDSS sample spans a range of $M_B = -26.1 \rm ~to~
-28.8$, and a redshift range of 3.6 to 5.0. Comparative numbers for
the APM sample observed by Omont et al. (1996a) are
$M_B = -26.8 \rm ~to~ -28.5$, and $4.0 \le z \le 4.7$. 

Our observations of the SDSS QSO sample include sensitive radio
continuum imaging at 1.4 GHz with the Very Large Array (VLA), and
photometry at 250 GHz using the Max-Planck Millimeter Bolometer array
(MAMBO) at the IRAM 30m telescope.  These observations are a factor
three more sensitive than previous studies of high redshift QSOs at
these wavelengths (Schmidt et al. 1995, Omont et al. 1996a), and are
adequate to detect emission powered by star formation in the host
galaxies of the QSOs at the level seen in low redshift ultra-luminous
infrared galaxies (L$_{FIR}$ $\ge$ 10$^{12}$ L$_\odot$; Sanders and
Mirabel 1999).  We will use these data to measure the correlations
between optical, mm, and cm continuum properties, and optical emission
line properties, and look for trends as a function of redshift. 

In this letter we present the first two mm detections from this study.
The sources are SDSSp J033829.31+002156.3 (hereafter SDSS 0338+0021),
and SDSSp J015048.83+004126.2 (hereafter SDSS 0150+0041); these
objects' names are their J2000 coordinates. They have $z = 5.00 \pm
0.04$ with $i^* = 19.96$, and $z = 3.67 \pm 0.02$ with $i^* =18.20$
(Fan et al. 1999). The absolute blue magnitudes of these quasars are
--26.56 and --27.75, respectively, assuming H$_o$ = 50 km s$^{-1}$
Mpc$^{-1}$ and q$_o$ = 0.5.   
These two sources  were taken from the Fall survey sample
(Fan et al. 1999), which covered a total area of 140 deg$^2$. 
Note that  J0150+0041 is the second most lumininous QSO in the 
combined Fall and Spring samples of Fan et al. (1999, 2000). 
 
\section{Observations and Results}

Observations  were made using MAMBO (Kreysa et al. 1999) at
the IRAM 30m telescope in December 1999 and February 2000. 
MAMBO is a 37 element
bolometer array sensitive between 190 and 315 GHz. 
The half-power sensitivity range is 210 to 290 GHz, and
the effective central frequency for a typical
dust emitting source at high redshift is 250 GHz.  
The beam for the feed horn of each bolometer is
matched to the telescope beam of 10.6$''$, and the bolometers are
arranged in an hexagonal pattern with a beam separation of
22$''$. Observations were made in standard on-off mode, with 2 Hz
chopping of the secondary by 50$''$ in azimuth.  
The data were reduced using
IRAM's NIC and MOPSI software packages (Zylka 1998).  
Pointing was monitored every hour, and was found to be
repeatable to within 2$''$. The sky opacity was monitored every
hour, with  zenith opacities between  0.23 and 0.36. Gain
calibration was performed using observations of Mars, Neptune, and
Uranus.  We estimate a 20$\%$ uncertainty in absolute flux
density calibration based on these observations.  The target sources
were centered on the central bolometer in the array (channel 1),
and the temporally correlated variations of the sky signal (sky-noise) 
detected in the remaining bolometers was subtracted from all the
bolometer signals. The total on-target plus off-target observing time 
for SDSS 0338+0021 was 108 min, while that for SDSS 0150+0041 was
77 min.  

The source SDSS 0338+0021 was detected with a flux density of
$3.7 \pm 0.5$mJy (Figure 1).  At $z = 5.0$, 250 GHz corresponds to an
emitted frequency of 1500 GHz, or a wavelength of 200 $\mu$m.
The source SDSS 0150+0041 was detected at 250 GHz
with a flux density of $2.0 \pm 0.4$ mJy (Figure 1). 
At $z = 3.7$, 250 GHz corresponds to an emitted frequency of 1180 GHz,
or a wavelength of 254 $\mu$m.  
The quoted errors in flux density do not
include the $20\%$ uncertainty in gain calibration. The sources 
were not seen to vary dramatically ($\le 30\%$)
between December 1999 and February 2000. 

Assuming a dust spectrum of the type seen in the low redshift starburst
galaxy Arp 220 (corresponding roughly to a modified black body
spectrum of index 1.5 and temperature of 50 K), 
the implied 60$\mu$m luminosity (Rowan-Robinson et al. 1997) 
for SDSS 0338+0021 is L$_{60} = 8.2 \pm 1.0  \times
10^{12}$ L$_\odot$, while that
for 0150+021 is  L$_{60} = 5.0 \pm 1.0 \times 10^{12}$ L$_\odot$.
Increasing the temperature to 100 K would increase the values of
L$_{60}$ by a factor of about 3.5, while using the spectrum of M82 as
a template would increase the values by a factor of 1.5. 

The 1.4 GHz VLA observations were made in the A 
(30 km) and BnA (mixed 30 km and 10 km) configurations
on July 8, August 14, September 30, and
October 8, 1999, using a total bandwidth of 100 MHz with two
orthogonal polarizations.  Each source was observed for a total of 2
hours, with short scans made over a large range in hour angle to
improve Fourier spacing coverage. Standard phase and amplitude
calibration were applied, as well as self-calibration using background
sources in the telescope beam. The absolute flux density scale was set 
using observations of 3C 48.  The final images were generated using
the wide field imaging and deconvolution capabilities of the AIPS task
IMAGR.  The observed rms noise values on the images were within
30$\%$ of the expected theoretical noise. The Gaussian restoring CLEAN 
beam was between 1.5$''$ and 2$''$ FWHM.

Neither source was detected at 1.4 GHz.  For SDSS 0338+0021 the
observed value at the source position was $37 \pm 24$ $\mu$Jy
beam$^{-1}$, and the peak in an 8$''$ box centered on the source
position was 60$\mu$Jy beam$^{-1}$ located 3.4$''$ southwest of the
source position.  At $z = 5.0$, 1.4 GHz corresponds to an emitted
frequency of 8.4 GHz.  For SDSS 0150+0041 the observed value at the
source position was $-3.0 \pm 19$ $\mu$Jy beam$^{-1}$, and the peak in
an 8$''$ box centered on the source position was 55$\mu$Jy beam$^{-1}$
located 2.8$''$ north of the source position.  At $z = 3.7$, 1.4 GHz
corresponds to an emitted frequency of 6.6 GHz.  We quote maximum and
minimum values in an 8$''$ box in the eventuality that the dust
emission is not centered on the QSO position (see discussion below).
In the analysis below we adopt an upper limit of 60$\mu$Jy beam$^{-1}$
for both sources.

Assuming a radio spectral index of --0.8, the 3$\sigma$ upper limit to
the rest frame spectral luminosity at 5 GHz of SDSS 0338+0021 is $3.3
\times 10^{31}$ erg s$^{-1}$ Hz$^{-1}$, while that for 0150+0041 is
$1.7 \times 10^{31}$ erg s$^{-1}$ Hz$^{-1}$.  These upper limits
place the sources in the
radio quiet regime, using the division at $10^{33}$ erg s$^{-1}$
Hz$^{-1}$ at 5 GHz suggested by Miller et al. (1990).  Note that
90$\%$ of optically selected 
high redshift QSOs are radio quiet according to this
definition (Schmidt et al. 1995).
 
\section{Discussion}

The fact that the cm flux densities for these sources are at least two
orders of magnitude below the mm flux densities, and that the sources
do not appear to be highly variable at 250 GHz, argues that the mm
signal is thermal emission from warm dust.  Adopting the relation
between L$_{60}$ and dust mass for hyper-luminous infrared galaxies
(i.e. galaxies with L$_{60}$ $\ge 10^{13}$ L$_\odot$), derived by
Rowan-Robinson (1999), the dust mass in SDSS 0338+0021 is $3.5 \times
10^8$ M$_\odot$, while that in SDSS 0150+0021 is $2 \times 10^8$
M$_\odot$.  This gives gas masses of $10 \times 10^{10}$ M$_\odot$,
and $6 \times 10^{10}$ M$_\odot$, respectively, using the
dust-to-molecular gas mass ratio of 300 suggested by Rowan-Robinson
(1999).  Of course, such estimates using data at a single frequency
are quite uncertain due to the lack of knowledge of
the dust temperature and emissivity law, 
and uncertainties in the gas-to-dust
ratio. We are currently searching for CO emission from these two
sources to obtain a better understanding of the gas content of these
QSOs.

Omont et al. (1996a) found that dust emission
seems to be correlated
with broad absorption line systems (BALs) in the APM QSO sample. 
The source SDSS 0150+0041 has been classified as a
`mini-BAL' by Fan et al. (1999). The source SDSS 0338+0021 
also shows strong and broad associated C IV absorption 
in high signal-to-noise spectra (Songaila et al. 1999). The
presence of strong associated absorption
is yet another indication of a rich gaseous
environment in these systems.  

The critical question when interpreting thermal dust emission from
high redshift QSOs is whether the emission is powered by the AGN or
star formation.  This question has been considered in great detail for
ultra- and hyper-luminous infrared galaxies (Sanders \& Mirabel 1996,
Sanders et al. 1989, Genzel et al. 1998), and a review of this
question can be found in Rowan-Robinson (1999).  Dust emitting
luminous QSOs, such
as SDSS 0338+0021 and SDSS 0150+0041, comprise a subset of the ultra-
and hyper-luminous infrared galaxies.  
In QSOs, typically less than 30$\%$ of the bolometric luminosity
is emitted in the infrared. Using the blue luminosity bolometric
correction factor of 16.5 derived for the PG QSO sample by
Sanders et al. (1989), the FIR luminosity of SDSS 0338+0021
accounts for only 15$\%$ of its bolometric luminosity,
and the FIR emission of SDSS 0150+0041 comprises only
3$\%$ of its bolometric luminosity. 
It is important to keep in mind that the FIR luminosity estimates
are based on measurements at a single frequency, and hence have
at least a factor two uncertainty (Adelberger \& Steidel 2000), 
while the bolometric luminosity estimates require an extrapolation
from the rest frame UV measurements into the blue, and hence have
comparable uncertainties.
Still, it is likely that only a minor fraction of
the total AGN luminosity needs to be absorbed and re-emitted
by dust to explain the FIR luminosity in both SDSS 0338+0021 and 
SDSS 0150+0041. 

An important related point is that the IR emitting region has a
minimum size of about 0.5 kpc for sources such as SDSS 0338+0021 and
SDSS 0150+0041, as set by the far IR luminosity and assuming optically
thick dust emission with a dust temperature of 50 K (Benford et
al. 1999, Carilli et al. 1999).  In the Sanders et al. (1989) model
for AGN-powered IR emission from QSOs, dust heating on kpc scales is
facilitated by assuming that the dust is distributed in a kpc-scale
warped disk, thereby allowing UV radiation from the AGN to illuminate
the outer regions of the disk. Detailed models by Andreani,
Franceschini, and Granato (1999) and Willott et al. (1999) show that
the  dust emission spectra from 3$\mu$m to 30$\mu$m can be 
explained by such a model. 

The alternative to dust heating by the AGN is to assume that there is
active star formation co-eval with the AGN in these systems.  Omont et
al. (1996) and Rowan-Robinson (1999) argue that star formation would
be a natural, although not required, consequence of the large gas
masses in these systems.  Imaging of a few high redshift dust emitting
AGN shows that in some cases the dust and CO emission come from
regions that are separated from the location of the optical AGN by
tens of kpc (Omont et al. 1999b, Papadopoulos et al. 2000, Carilli et
al. 2000).  Such a morphology argues strongly for a dust heating
mechanism unrelated to UV emission from the AGN, at least in these
sources. Lastly, Rowan-Robinson (1999) performed a detailed analysis
of the spectral energy distributions (SEDs) of hyper-luminous infrared
galaxies and concluded that, although $\ge 50\%$ of these systems show
evidence for an AGN in the optical spectra, the SEDs between 50$\mu$m
and 1mm are best explained by dust heated by star formation. Yun et
al. (2000) have extended this argument to cm wavelengths and reach a
similar conclusion. 

Carilli \& Yun (2000) present a model for the expected behavior with
redshift of the observed spectral index between cm and submm
wavelengths for star forming galaxies, relying on the tight
radio-to-far IR correlation found for nearby star forming galaxies
(Condon 1992).  
Using the observed mm
flux densities of SDSS 0150+0041 and SDSS 0338+0021,  
these models predict  flux densities
at 1.4 GHz between 15 and 60 $\mu$Jy for both sources. 
Radio images with a factor three better sensitivity are required to
determine if these two sources have cm-to-mm SEDs
consistent with low redshift star forming galaxies. 

If the dust and radio continuum emission is powered by star formation
in SDSS 0338+0021 and SDSS 0150+0041, then the star formation rates in
these galaxies are high.  Using the relation between L$_{60}$ and
total star formation rate given in Rowan-Robinson (1999) for a $10^8$
year starburst assuming a standard Salpeter IMF, the implied rate for
SDSS 0338+0021 is 2700 M$_\odot$ year$^{-1}$, while that for SDSS
0150+0041 is 1800 M$_\odot$ year$^{-1}$.  Note that the high redshift
QSOs from the APM sample were detected at flux levels between 3mJy
and 12mJy at 240 GHz (Omont et al. 1996a), hence the required star formation
rates may be even larger in some systems. The implication is that we may
be witnessing the formation of a large fraction of the stars of the
AGN host galaxy on a timescale $\le 10^8$ years.  

Magnification by
gravitational lensing would lower the required luminosities, bringing
the sources more in-line with known ultra-luminous infrared galaxies.
Two of the APM QSOs detected at 240 GHz by Omont et al. (1996a)
show possible evidence for gravitational lensing, while 
two other high redshift dust emitting QSOs, H1413+117 and APM
08279+5255, are known to be gravitationally lensed (Barvainis et
al. 1994, Downes et al. 1999).  However, thus far there is
no evidence for multiple imaging on arc-second scales for either SDSS
0338+0021 or SDSS 0150+0041 in optical and near IR images (Fan et al.
2000).  Imaging at sub-arcsecond resolution is required to determine
if these sources are gravitationally lensed.

An important question concerning the formation of objects in the
universe is: which came first, black holes or stars (Rocca-Volmerange
et al. 1993)?  If the dust emission is powered by a starburst in SDSS
0338+0021 and SDSS 0150+0041, then the answer to the above question in
some systems may be: both.  Co-eval starbursts and AGN at high
redshift may not be surprising, since both may occur in violent galaxy
mergers as predicted in models of structure formation via hierarchical
clustering (Franceschini et al. 1999, Taniguchi, Ikeuchi, \& Shioya
1999, Blain et al. 1999,  Kauffmann \& Haehnelt 2000, Granato et
al. 2000).    To properly address the interesting question of 
AGN versus starburst dust heating in these high redshift QSOs
requires well sampled SEDs from  cm to optical wavelengths, 
and perhaps most importantly, imaging of the mm and cm continuum, and
CO emission, with sub-arcsecond resolution. 

\vskip 0.2truein 

The VLA is a facility of the National Radio
Astronomy Observatory (NRAO), which is operated by Associated
Universities, Inc. under a cooperative agreement with the National
Science Foundation.
This work was based on observations carried out with the IRAM 30 m
telescope.  IRAM is supported by INSU/CNRS (France), MPG (Germany) and
IGN (Spain).  This research made use of the NASA/IPAC Extragalactic
Data Base (NED) which is operated by the Jet propulsion Lab, Caltech,
under contract with NASA.  CC acknowledges support from the Alexander
von Humboldt Society.  DPS acknowledges support from National Science
Foundation Grant AST99-00703.  XF and MAS acknowledge support from the
Research Corporation, NSF grant AST96-16901, and an Advisory Council
Scholarship.


\newpage

\begin{figure}
\caption{The time averaged, opacity corrected, sky-noise subtracted
flux densities derived for 35 of the 37 bolometers (two channels
were disfunctional). 
The thick line represents the time-averaged signal from the central, 
on-source bolometer (channel 1). The thin solid line is the 
mean of the off-source bolometers, and 
the dashed lines 
show the 1$\sigma$ noise envelope for the off-source bolometers. 
The upper figure shows observations of SDSS 0338+0021 and the lower
figure shows observations of SDSS 0150+0041.
}
\end{figure}

\begin{figure}
\vskip -0.5in
\psfig{figure=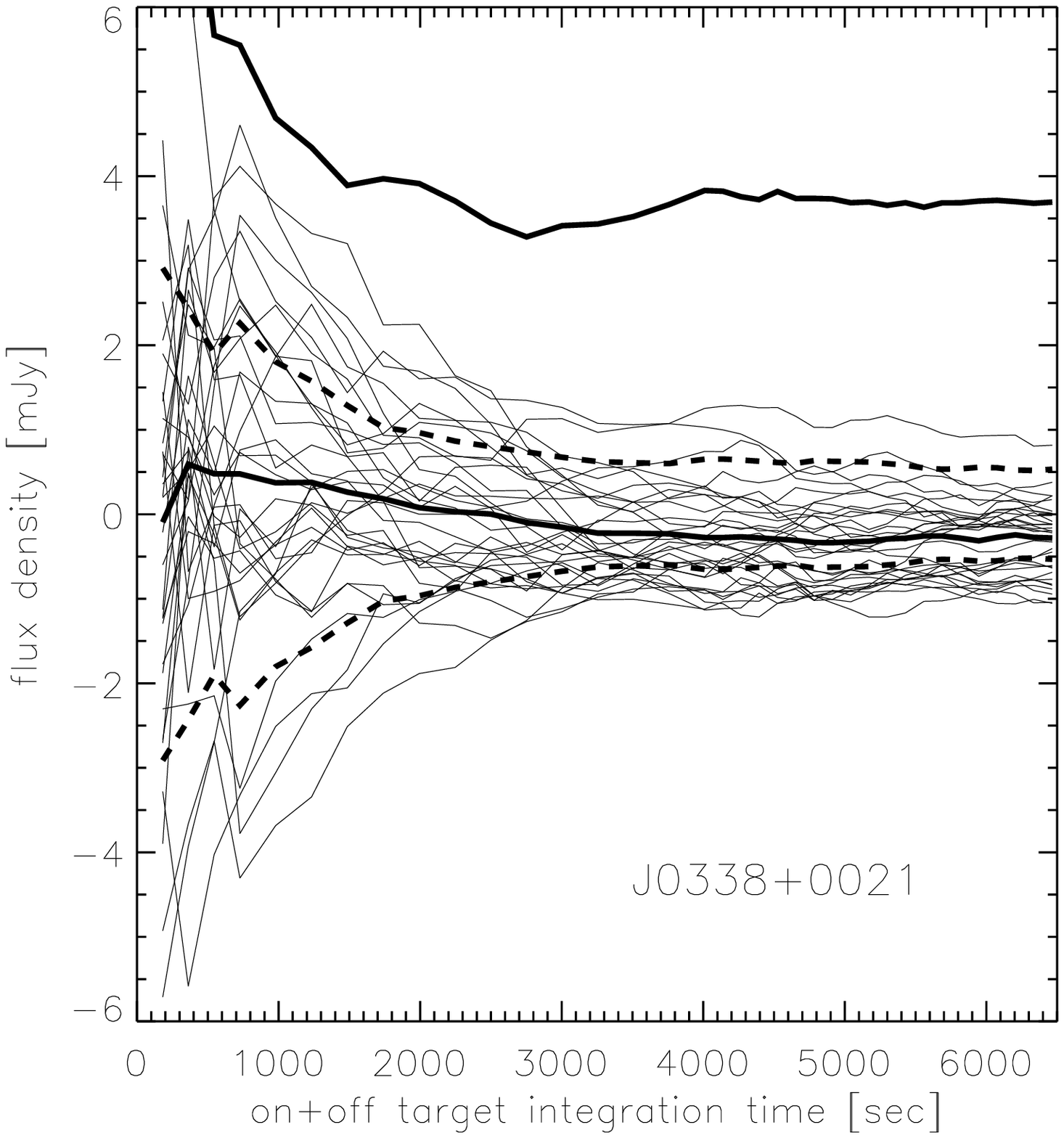,width=4in}
\psfig{figure=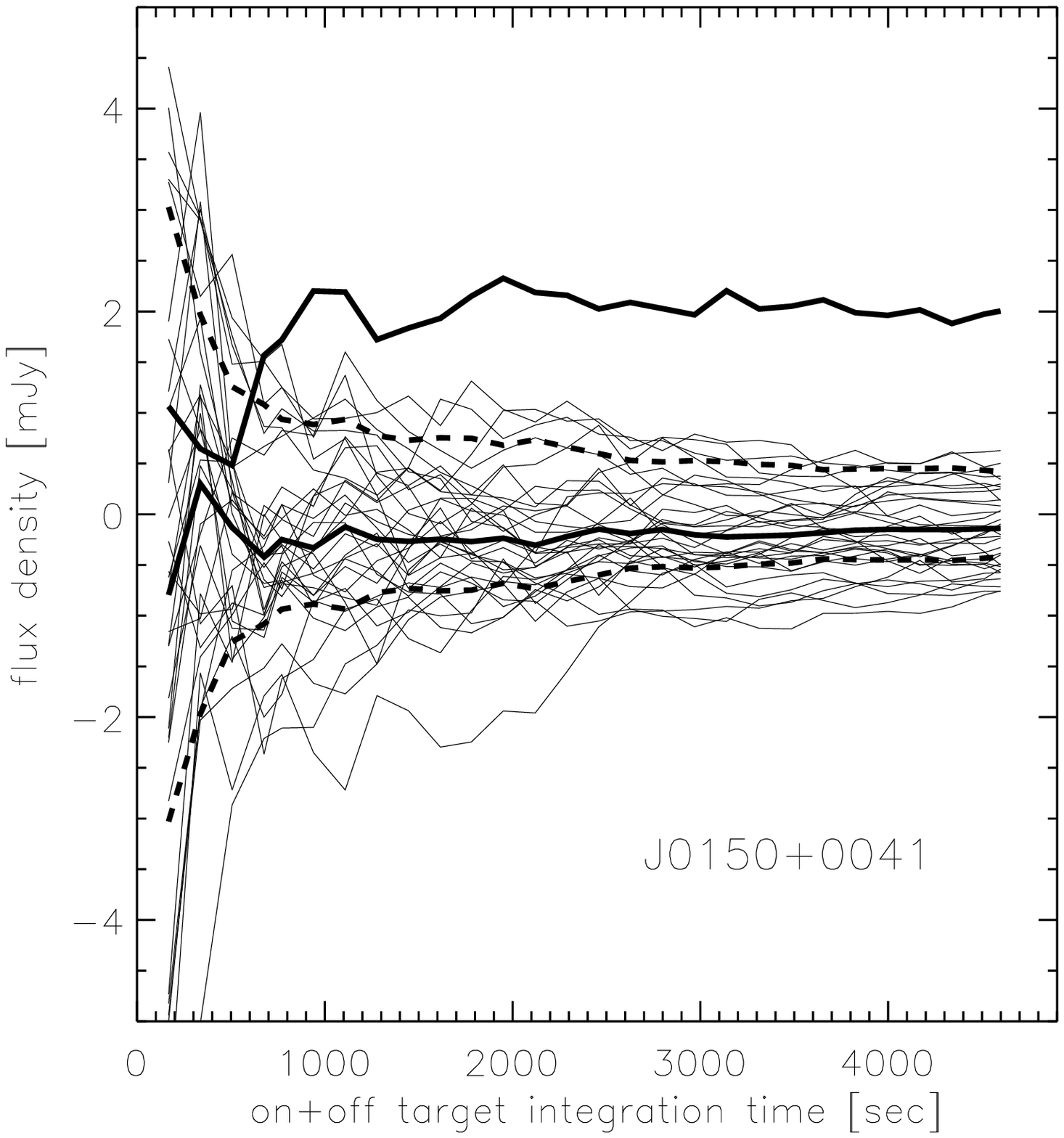,width=4in}
\end{figure}

\vfill\eject

\end{document}